\documentclass{elsart}

\usepackage{psfig}
\usepackage{amssymb}
\journal{Physica D}
\begin{document}

\begin{frontmatter}



\title{Travelling fronts in an array of coupled 
symmetric bistable units}


\author{Diego Paz\'o}\footnote{Tel.: +34-981-547023;
fax: +34-981-522089; e-mail: diego@fmmeteo.usc.es; http://chaos.usc.es}
\author{and Vicente P\'erez-Mu\~nuzuri}

\address{Grupo de F\'{\i}sica non Lineal, Facultade de F\'{\i}sica,\\
Universidade de Santiago de Compostela,
E-15782 Santiago de Compostela, Spain}

\begin{abstract}
A symmetry breaking mechanism is shown to occur in an
array composed of symmetric bistable Lorenz units
coupled through a nearest neighbour scheme.
When the coupling is increased,
we observe the route: {\em standing} 
$\rightarrow$ {\em oscillating} 
$\rightarrow$ {\em travelling} front. In some circumstances,
this route can be described in terms of a gluing of two 
cycles on the plane. In this case, the asymptotic 
behaviour of the velocity of the front is
found straightforward. 
However, it may also happen that the gluing bifurcation involves 
a saddle-focus fixed point. 
If so, front dynamics may become quite complex displaying
several oscillating and propagating regimes, including chaotic
(Shil'nikov-type) front propagation. Finally, we compare our
results with different couplings and with other discrete 
bistable media. 
\end{abstract}

\end{frontmatter}

% main text

\section{Introduction}
\label{intro}

Several areas of science use model equations of the
reaction-diffusion type. Moreover, besides its
special interest in some fields (chemistry~\cite{turing,fife}, 
biology~\cite{murray}, ...) 
reaction-diffusion equations
are considered as abstract models for pattern 
formation~\cite{cross}.

Real systems are frequently composed of discrete elements,
material models and biological cells are just two examples.
Therefore, it is natural to deal with the discrete space 
version of the reaction-diffusion equation. 
In this point, it must be emphasized that discreteness 
may manifest itself in a strong way, such that some 
phenomena are not simple analogies of the continuous ones.

Here, we focus on a discrete reaction-diffusion
model such that its local dynamics (reaction term) is bistable,
with two stable fixed points. 
Particularly important examples of bistable dynamics are found 
in optics~\cite{bowden,firth}, chemical systems~\cite{laplante}, and 
biology~\cite{murray,mckean,fitzhugh}. The main interest 
of these systems lies in the behaviour of the fronts that 
constitute the boundaries of the domains of both (stable) states.
Intuitively, the front will move from the most to
the less stable state, enlarging a domain and shrinking 
the other. In fact, great attention has been devoted to the phenomenon
of `propagation failure' or 
`pinning'~\cite{keener,erneux,vicente,mackay,kladko,carpio}, 
because it is usual that in discrete systems some threshold
must be surpassed to achieve propagation.

But, although one could expect that propagation does not succeed
in the case of symmetric bistability, it was demonstrated
that such possibility exists.
Hagberg and Meron~\cite{meron} studied a continuous bistable 
reaction-diffusion system, the FitzHugh-Nagumo model, showing
that, in the symmetric case, propagation occurred after a symmetry
breaking front bifurcation. In this scenario, the front breaks its 
symmetry through a pitchfork bifurcation, at the same time 
that propagation is initiated. Also, an analogous bifurcation 
was found for the complex Ginzburg-Landau equation~\cite{coullet} 
with the name of nonequilibrium Ising-Bloch bifurcation. 

In a recent work, we have reported~\cite{predic} the existence of 
front propagation in a discrete symmetric bistable medium.
The transition is exclusive for a discrete system and it is possible
thanks to the multi-variable nature of the local 
dynamics. One variable
systems are not able to break symmetry~\cite{keener} even if
the function that describes the local dynamics is not 
continuous~\cite{fath}.
It was shown that the mechanism leading to propagation is not 
a pitchfork bifurcation. Contrastingly, it consists on a 
Hopf bifurcation followed by a global bifurcation, 
that is equivalent to a gluing bifurcation of cycles. 
In correspondence, 
the velocity of the front shows a logarithmic dependence
close to the onset. The aim of the present contribution 
is two-fold. On one hand, a more detailed analysis
of the transition is undertaken, explaining which
is the role of the stationary states of the system.
On the other hand, the new features of the transition
when a saddle-focus fixed point mediates the 
gluing bifurcation are investigated.

The paper is organized as follows: Section~\ref{system} 
presents the model. In Sec.~3 an overview of the 
transition is presented. In Sec.~4 a reduced 
cylindrical coordinate system is defined, and the
mechanism of gluing of cycles underlying the
transition between oscillating and travelling
fronts is explained. In Sec.~5, several 
non-trivial front dynamics are shown, including
the chaotic motion of the front. Finally, Sec.~6 is devoted to
further considerations, while in Sec.~7 we present the
conclusions. 

\section{The system}
\label{system}

We consider a discrete reaction-diffusion equation in
one dimension:
\begin{equation}
\dot{{\bf r}}_j= f({\bf r}_j)+
{{D}\over{2}}\Gamma ({\bf r}_{j+1}+{\bf r}_{j-1}-2{\bf r}_j) ,
\label{general}
\end{equation}
where $D$ accounts for the coupling strength between neighbours, and
the coupling matrix $\Gamma$ says which variables get coupled.
In what concerns the local bistable dynamics, 
we deal with the well-known Lorenz oscillator~\cite{lorenz}
as unit cell. It exhibits several characteristic 
behaviours depending on its internal parameters~\cite{sparrow}: 
monostability,
bistability, limit cycle, `butterfly chaos', `noisy periodicity', etc.
We consider the range where the system contains two stable 
symmetry related fixed 
points (bistability). The coupling matrix $\Gamma$ is taken with
all elements zero except one: $\Gamma=\gamma_{kl}=\delta_{k2}\delta_{l2}$.
Therefore, the equations describing the temporal evolution of the array are:

\begin{eqnarray}
\dot{x}_j&=&\sigma(y_j-x_j) \nonumber\\
\dot{y}_j&=&r x_j-x_j z_j - y_j + {D\over2} (y_{j+1}+y_{j-1}-2y_j) 
 \qquad\qquad j=1,\ldots,N \label{lorenz}\\
\dot{z}_j&=&x_j y_j- b z_j \nonumber
\end{eqnarray}

The standard values are chosen for the $\sigma$ and $b$ 
parameters: $\sigma=10$ and $b=8/3$. Then, the Lorenz system
contains two stable foci 
$C_{\pm}=$ $(\pm\sqrt{b (r-1)},$ $\pm\sqrt{b (r-1)},$ $r-1)$
for $r$ in the 
range $(1.35,r_H)$; at $r=r_H \approx 24.74$, 
$C_{\pm}$ become unstable through a subcritical Hopf bifurcation. 
Also, it is important to note that 
there exists a saddle fixed point located at the origin
for $r>1$.

A fourth-order Runge-Kutta method, with time step 0.01, was used
to integrate Eq.~(\ref{lorenz}). A step-like initial condition
is imposed to the system.
Results on front propagation do not depend on the 
boundary conditions provided that the array is large enough.
Thus, if one wishes the system evolve for long time 
in the regime of travelling front, one 
should move the boundary whilst the front propagates.
Also, one may imposed to
the system periodic boundary conditions, with
{\it two} fronts diametrically opposed as initial
conditions. If the initial condition satisfies $x_j=-x_{j+N/2}$,
$y_j=-y_{j+N/2}$, $z_j=z_{j+N/2}$,
both fronts propagate always in the same direction,
avoiding its mutual annihilation.  

%%%%%%%%%%%%%%%%%%%%%%%%%%%%%%%%%%%%%%%%%
\section{Standing, oscillating and travelling fronts}
\label{propag}

\begin{figure}
\begin{center}
\psfig{file=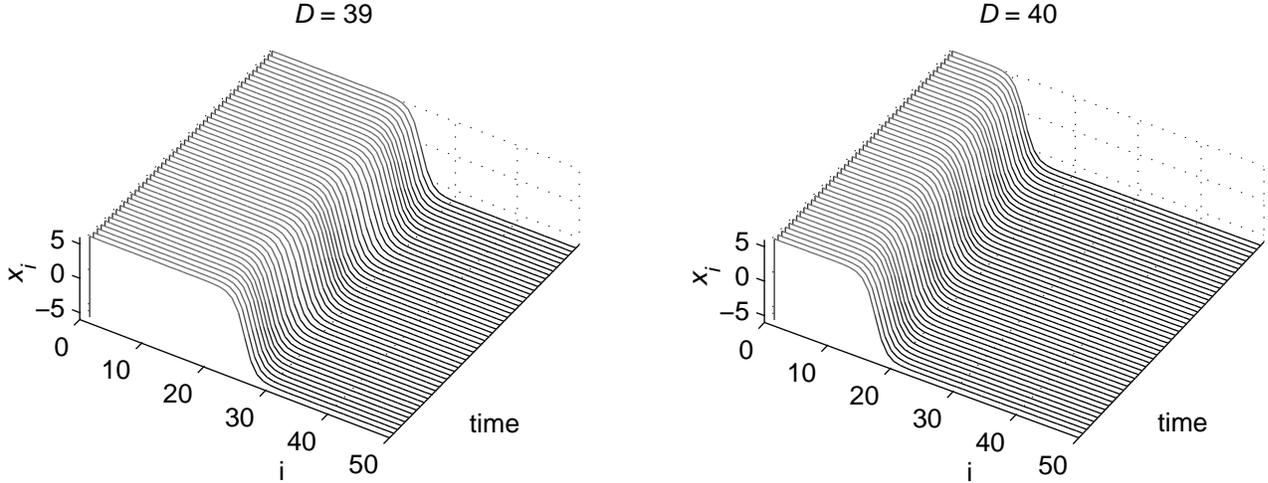}
\caption[]{Spatio-temporal evolution of the front for two
different values of the coupling strength $D$ and $r=14$. 
A step-like initial condition is imposed for an open array 
of $N=50$ oscillators. When the coupling surpasses the critical 
value $D=D_{th}\approx39.63$ the front propagates
through the array shifting the oscillators from $C_+$ to $C_-$.
The time interval shown is 100 time units.}\label{3d} 
\end{center} \end{figure}

Three main front dynamics are found when 
varying the parameter $r$ ($r<r_H$), and the coupling
strength $D$ ($D>0$). We distinguish among standing (static), 
oscillating and travelling fronts. In Fig.~\ref{3d}, a 
centered step-like 
initial condition is imposed to the system with free ends
and $r=14$; two different 
non-static regimes are achieved depending on the value of the 
coupling strength $D$. Note
that the Lorenz model is symmetric and the coupling that
appears in Eq.~(\ref{lorenz}) destroys this local symmetry
but preserves the global reflection (or ${\bf Z}_2$) symmetry:

\begin{equation}
(x_1, y_1, z_1, ..., x_N, y_N, z_N) \longrightarrow  
(-x_1, -y_1, z_1, ..., -x_N, -y_N, z_N)
\label{z2}
\end{equation}

\begin{figure}
\begin{center}
\psfig{file=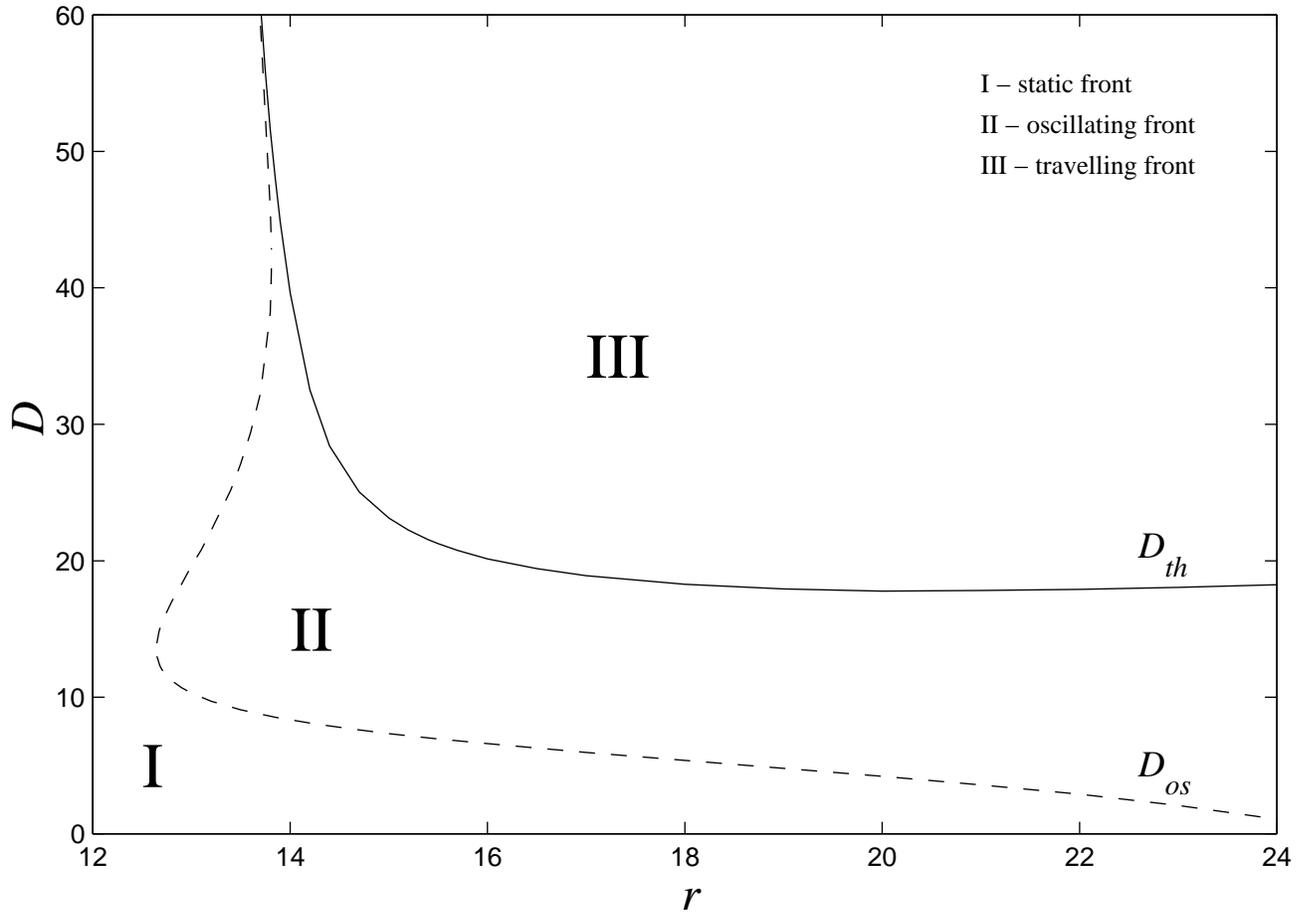}
\caption[]{Three main types of dynamics are distinguished
as a function of $r$ and $D$. The boundaries among them
are the critical lines $D_{os}$ and $D_{th}$.}\label{diagrama}
\end{center} \end{figure}

Therefore, for $N$ large enough, both senses of propagation 
are equally likely.
A diagram of the different regions
on the $r-D$ plane is shown in Fig.~\ref{diagrama}.
It may be seen that static fronts are found for $D$ and/or $r$
small, whereas travelling fronts appear for large $D$ and $r$.
Also, it is significative that oscillating fronts always exist
between the standing and the travelling fronts regions. 
The line ($D_{os}$), that defines the boundary between
the standing and the oscillating regions, approach $D=0$ as 
$r\rightarrow r_H$, since the nature of the fixed points
($C_{\pm}$) deeply influences the dynamics of the front. This is
not very surprising, but indicates that oscillating fronts
will be usually found in bistable systems with stable foci,
rather than nodes. Somehow, the stability properties 
of the fixed points are transmitted to the front.
Also, it is significative that $D_{th}$ 
(the threshold for travelling fronts) seems to diverge
asymptotically at $r = r_{\infty} \approx 13.5$. 

\begin{figure}
\begin{center}
\psfig{file=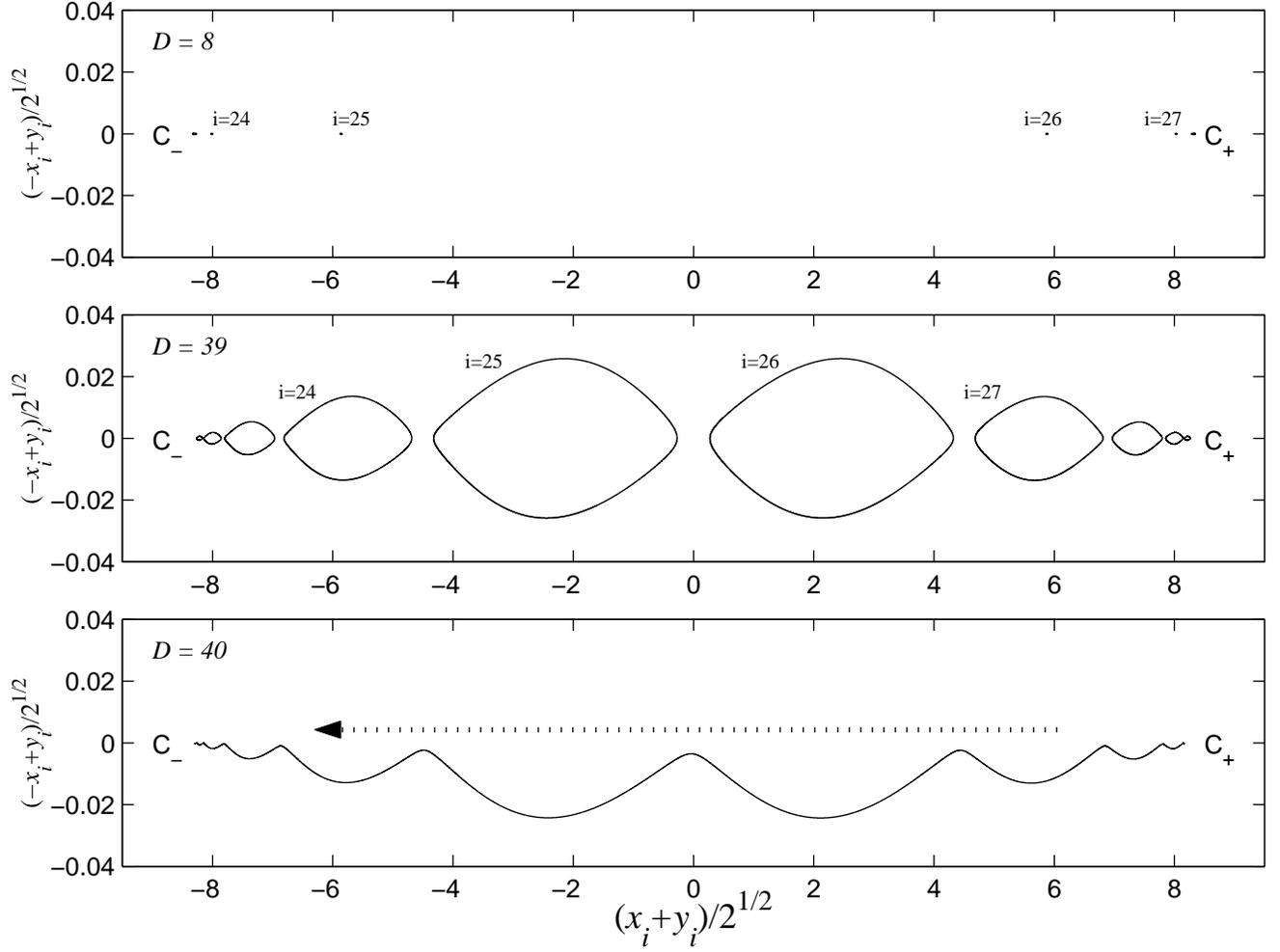}
\caption[]{Projections onto the reference system
($(x_i+y_i)/\sqrt{2}$, $(-x_i+y_i)/\sqrt{2}$)
of all the oscillators of the array with step-like configuration
and $r=14$. 
From top to bottom, standing, oscillating, and travelling cases are 
shown. In the travelling case, the line is the trajectory followed by 
each oscillator going from $C_+$ to $C_-$ as the front advances.} 
\label{multiple}
\end{center} 
\end{figure}

Instead of visualizing the system as in Fig.~\ref{3d}, we
show now which is the behaviour of the different units
when the coupling is increased from zero. It is clear that
for $D=0$ the step-like solution 
that we were considering
consists of ${\bf r}_j=C_+$, $j=1,\ldots,25$ and 
${\bf r}_j =C_-$, $j=26,\ldots,50$ ($N=50$). As the coupling 
increases this solution can be smoothly continued. 
But there exist two constrains for the stationary solutions, 
that come from Eq.~(\ref{lorenz}):
\begin{equation}
x_j=y_j , \qquad z_j={{x_j y_j}\over b} \qquad \forall j \in \{1, 2, ..., N\}
\label{fijos}
\end{equation}
Therefore, all the units lie on a parabola that 
passes through $C_{\pm}$ and zero. However, when
$D_{os}$ is reached the front undergoes a Hopf
bifurcation, and all the units start to oscillate
leaving the mentioned parabola. The projection of the
parabola onto the plane $(x,y)$ is a straight line, the bisectrix
of the first and third quadrants. However, to better visualize
oscillations out of the parabola, in Fig.~\ref{multiple}
we performed a $45^\circ$ rotation of the reference system.
Obviously those units
closer to the front oscillate with larger amplitude that
those located far from the front that are almost
insensitive to the bifurcation.
When $D_{th}$ is reached
a multiple collision occurs; the orbits of
neighbouring units collide creating two ``channels"
going from $C_{\pm}$ to $C_{\mp}$. Like in the
symmetric FitzHugh-Nagumo model studied by Hagberg and 
Meron~\cite{meron}, the travelling front is not symmetric. 
However, in our case the mechanism leading to propagation 
is not a pitchfork bifurcation, instead, a Hopf bifurcation
that creates the oscillating
front is the precursor of the travelling front.

\begin{figure}
\begin{center}
\psfig{file=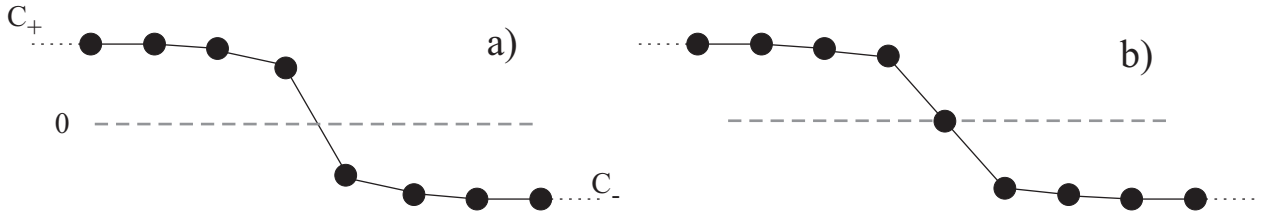,clip=true} 
\caption[]{A sketch of the two monotonically decreasing stationary
solutions for $D>0$. a) $A$-state, continuation of the 
state $(\ldots, C_+, C_+, C_-, C_-, \ldots)$ at $D=0$. b) $B$-state,
continuation of the 
state $(\ldots, C_+, C_+, {\bf 0}, C_-, C_-, \ldots)$ at $D=0$.} 
\label{2s}
\end{center} \end{figure}

The description provided by Fig.~\ref{multiple} is illustrative,
but it is incomplete if one does not realize that 
besides the static solution that loses its stability
through a Hopf bifurcation, there must exist another
(unstable) stationary state that mediates the multiple collision 
of cycles. Therefore, one must search the stationary solutions
monotonic in $\{x_j\}$ and $\{y_j\}$. It
is not difficult to find out that only two monotonic
solutions, called stable and unstable dislocations, exist (discarding 
spatial translations). They are the 
continuation of the solutions in the uncoupled limit:
$(\ldots, C_+, C_+, C_-, C_-, \ldots)$ and
$(\ldots, C_+, C_+, {\bf 0}, C_-, C_-, \ldots)$. 
We shall refer to them as $A$- and $B$-state,
respectively, and Fig.~\ref{2s} shows a sketch of them. It is the $B$-state
which mediates the transition at $D_{th}$, and it will be 
shown below that the stability properties of this solution
will determine important features of the onset of the 
wavefronts. 
%%%%%%%%%%%%%%%
\section{Gluing of cycles}

In this section we demonstrate how the transitions
presented in the previous section may be described
in the context of a gluing bifurcation, 
in which two limit cycles become a two-lobed cycle
by involving an intermediate saddle point.
The gluing bifurcations is usual in systems with ${\bf Z}_2$
symmetry~\cite{lopez,herrero}, because provided the existence
of the symmetry it becomes a codimension-one bifurcation.

\subsection{Cylindrical coordinates}

Figure \ref{multiple} shows some kind of collective motion when
the front oscillates. Then, the whole set of variables 
$(x_j, y_j, z_j)$ is discarded, and instead, we chose
to build a reduced phase
space to study the dynamics. Notice that if the medium is
infinite, there exists a perfect symmetry under translation
along the array. With this in mind, we define two auxiliary 
variables $\xi$ and $\eta$, that have sense for large enough $N$ and
when the front is far enough from the boundaries: 

\begin{eqnarray}
\xi &=& {1\over \sqrt{2}} \sum_{j=1}^{N} x_j+y_j  \qquad \bmod\left(2\sqrt{2 b (r-1)}\right)\\
\eta&=& {1\over \sqrt{2}} \sum_{j=1}^{N} -x_j+y_j .
\end{eqnarray}

Variable $\xi$ accounts for the propagation of the front and
is not bounded by default, then it must be
defined within the range $\Delta \xi =
[ 2 \sqrt{ b (r-1)}+ 2 \sqrt{ b(r-1)} ]/\sqrt{2} = 
2 \sqrt{2 b (r-1)}$, that is the magnitude that
$\xi$ increases or decreases when the front 
moves one position along the array. On the other hand, $\eta$ is defined so
that it is bounded under front propagation, notice that it is 
defined from Eq.~(\ref{fijos}).
If the medium consists of a number $N$ even of units, the
static solutions are located in this cyclic phase space in:

\begin{eqnarray}
A: \xi &=& \eta=0\\
B: \xi &=& \pm \sqrt{2 b (r-1)}, \eta=0 
\end{eqnarray} 

If $N$ is odd both solutions exchange their coordinates. In
what follows, we only take $N$ even, unless it is specified.

\begin{figure}
\begin{center}
\psfig{file=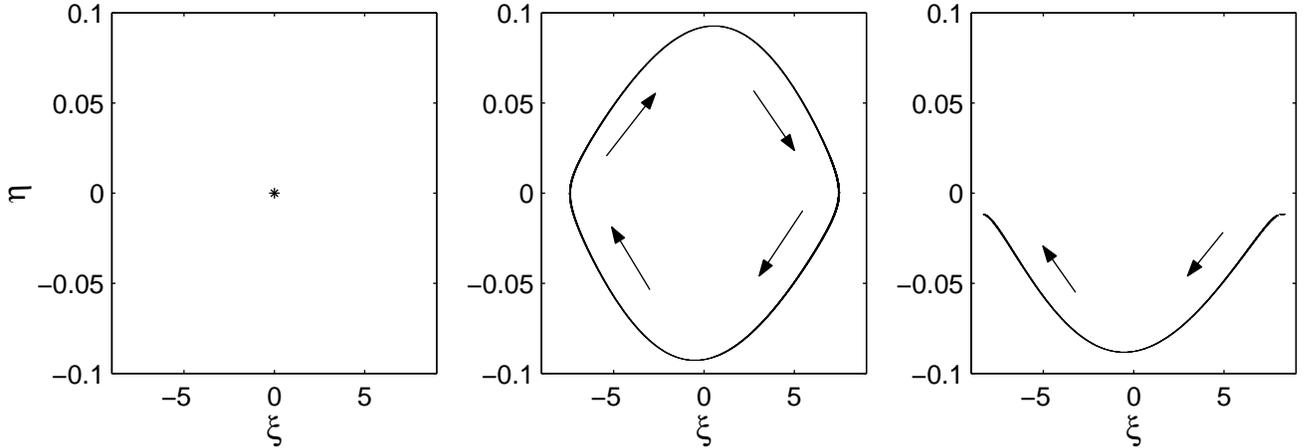}
\caption[]{Evolution on the reduced cylindrical 
phase space spanned by $\xi$ and $\eta$ for three different 
values of $D$ and
$r=14$; from left to right: standing, oscillating, and
travelling front. At $D=D_{th}\approx 39.63$ a double homoclinic
connection to the $B$-state arises.} 
\label{pendulo}
\end{center} 
\end{figure}

The transition shown in Fig.~\ref{multiple} 
is now shown in the cylindrical phase space in 
Fig.~\ref{pendulo}. Notice that when the
front is static, we have an equilibrium point at $\xi=\eta=0$
($A$-state). 
When the front oscillates a limit cycle exists, and finally
when the coupling goes beyond $D_{th}$ we find two
symmetry related limit cycles, that turn around 
the cylindrical phase space in opposite directions. 
The situation is quite similar to a pendulum
where libration corresponds now to oscillation, and
rotation of the pendulum corresponds to propagation.
Thus, at $D=D_{th}$ there exist two homoclinic loops that
connect the $B$-state with itself (called separatrices
for the pendulum). This bifurcation is called gluing
bifurcation because it involves the collision of two cycles
to create another; although in our case it could 
be more appropriate to talk of a inverse gluing 
or a splitting bifurcation.

Nonetheless, it is better to visualize
our gluing bifurcation in ${\mathbb R}^2$. A possible transformation
consists, topologically, in what follows:
First of all, cut the cylinder with two planes perpendicular
to its axis in such a way that the saddle point and the
heteroclinic orbits stand between both planes. Then compress
both circumferences that limit the ``cylinder" inbetween 
to a point. Thus, we have got an object topologically equal to a sphere.
These two steps could be substituted by making
the section of the cylinder equal to zero at $\eta=\pm \infty$
and then making a transformation that makes our object finite.
The last step is to make a hole at $\xi=\eta=0$ and deform what
remains into a plane.

\begin{figure}
\begin{center}
\psfig{file=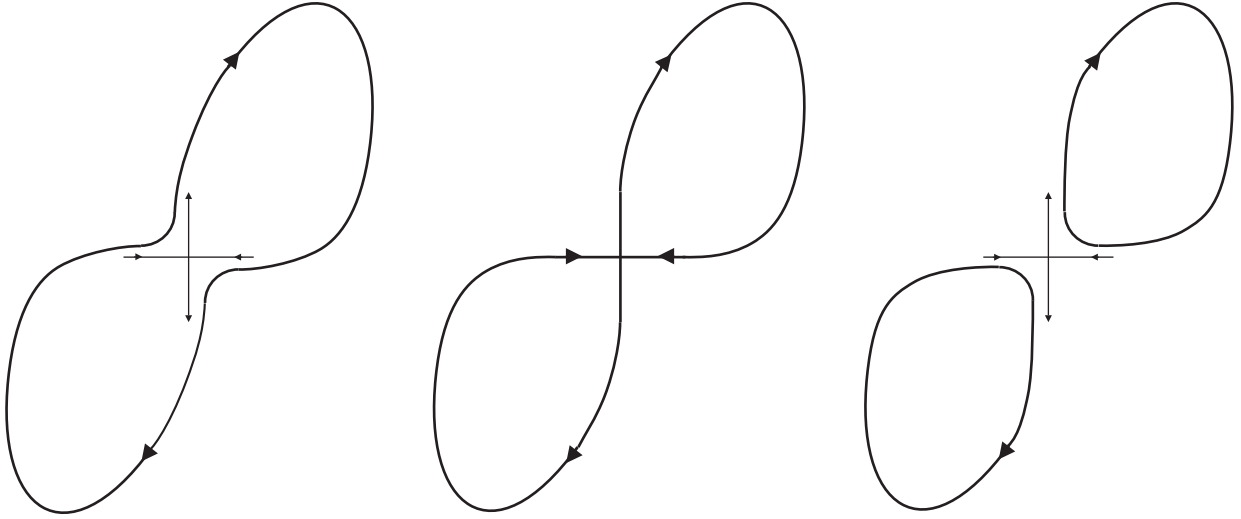}
\caption[]{Schematic of a (inverse) gluing bifurcation on $\Rset^2$. 
For $D<D_{th}$ the trajectory approaches twice per cycle the saddle 
point and the front oscillates. Beyond the critical value $D=D_{th}$
two symmetry related orbits, corresponding
to both senses of propagation of the front, appear. 
$\lambda_u$ and $\lambda_s$ stand
for the eigenvalues of the saddle fixed point
at homoclinicity.}
\label{gluing} 
\end{center} 
\end{figure}

In Fig.~\ref{gluing} the sketch of a 
gluing bifurcation equivalent to the one shown in 
the cylindrical phase space (Fig.~\ref{pendulo}) is depicted.
Notice that for $D<D_{th}$ there exist an orbit that 
approaches twice per period close to the saddle point
(the $B$-solution), whereas for $D>D_{th}$ two
symmetry related cycles, corresponding to both senses of
propagation, coexist.

\subsection{Velocity of the front as a function of $D-D_{th}$}

\begin{figure}
\begin{center}
\psfig{file=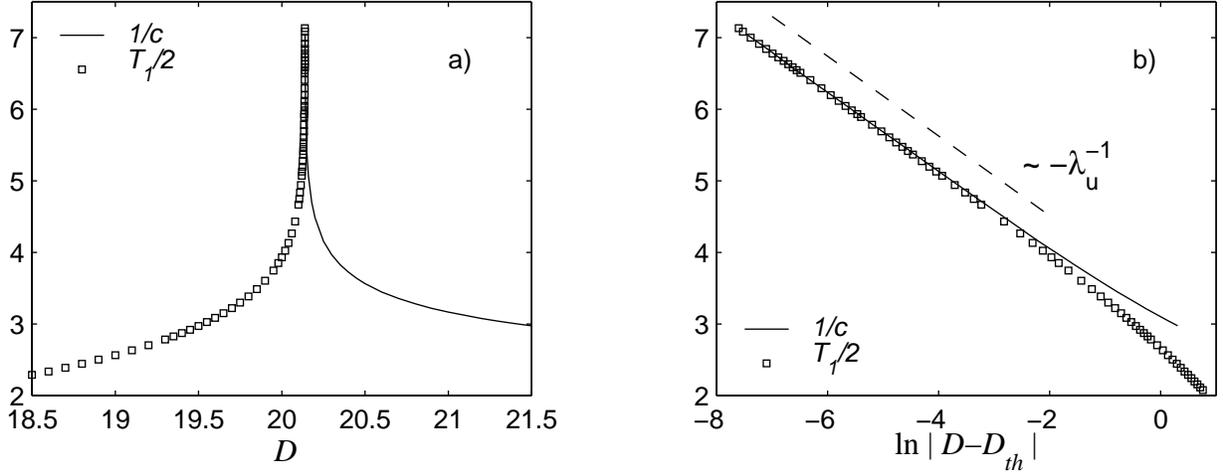}
\caption[]{$1/c$ (solid line) and $T_1/2$ (squares) as a 
function of $D-D_{th}$ (a) and $\ln(|D-D_{th}|)$ (b) for
$r=16$.
The behaviour agrees with that predicted by Eqs.~(\ref{t1}) and
(\ref{t2}). The critical coupling is found to be
$D_{th}^{r=16}=20.13952$. In (b) a straight line
with slope $-\lambda_u^{-1}$ is depicted, being 
$\lambda_u=1.7959$ the unstable eigenvalue of the $B$-state
computed numerically for an array of $N=21$ units.}
\label{r16} 
\end{center} 
\end{figure}

The collision (and subsequent destruction) of a periodic orbit
with a saddle, 
called saddle-loop bifurcation, is characterized
by a logarithmic lengthening of the period 
of the cycle~\cite{books}. For the scenario 
depicted in Fig.~\ref{gluing}, we expect to find
the following dependencies for the periods ($T_{1,2}$) 
in a neighbourhood of $D_{th}$:

\begin{eqnarray}
T_1&=&a_1 - {2\over\lambda_u} \ln (D_{th}-D) \label{t1} \\
\frac{1} {c}=T_2&=&a_2 - {1\over\lambda_u} \ln (D-D_{th}) ,
\label{t2}
\end{eqnarray} 
where $\lambda_u$ is the unstable eigenvalue of
the saddle fixed point.
Taking into account that one turn around the cylinder is
equivalent to the movement of the front in one cell, it is
clear that $T_2$ is the inverse of the velocity of the front ($c$).
In the equation for the period $T_1$ a factor 2 appears because
the orbit approaches twice per cycle to the neighbourhood of
the saddle point. Moreover, it is expected that the `fast
dynamics' (the motion far from the saddle) contained in variables 
$a_{1,2}$ will be approximately
the same at both sides of the transition, and then, $a_1\approx 2 a_2$. 
In Ref.~\cite{predic} it was shown that at $r=14$, the 
behaviour of $T_1/2$ and $c^{-1}$ agrees with the
predictions given by Eqs.~(\ref{t1},\ref{t2}), and we show it here 
for $r=16$ in Fig.~\ref{r16}.

Notice that the velocity of the front grows quite
abruptly from zero at $D=D_{th}$. The derivative of
$c=c(D)$, near $D_{th}$ is, according to Eq.~(\ref{t2}):

\begin{equation}
\def\mapright#1{\smash{\mathop{\longrightarrow}\limits^{#1}}}
\frac{dc}{dD}=\frac{1}{\lambda_u (D-D_{th})} 
\left ( \frac{1}{a_2- \frac{1}{\lambda_u}\ln (D-D_{th})} \right )^2
\qquad \mapright{D \rightarrow D_{th}} \infty  .
\end{equation}

\subsection{Quantitative analysis}

In order to verify, not only qualitatively, but quantitatively
the tendency of $T_1$ and $c$ when $D \rightarrow D_{th}$ 
we computed numerically the value of the eigenvalues of
the $B$-state at $D=D_{th}$ for finite $N$. 
To preserve the symmetry, when computing the eigenvalues,
the number of units $N$ was chosen to be odd, with the central unit
located at the origin.
The jacobian matrix ($J$) for a discrete reaction-diffusion
model (see Eq.~(\ref{general}))
exhibits the following tridiagonal structure:

\begin{equation}
 J= \,
\left ( \begin{array}{cccccc} 
{\bf H}_1  & {\bf D}    & {\bf 0}   & \cdots  & {\bf 0}  & {\bf 0}\\
 {\bf D}   & {\bf H}_2  & {\bf D}   & \cdots & {\bf 0}  & {\bf 0}\\
{\bf 0}    & {\bf D}    & {\bf H}_3 & \cdots  & {\bf 0}  & {\bf 0}\\
\vdots     &   \vdots   & \vdots    & \ddots  & \vdots   & \vdots \\
{\bf 0}    & {\bf 0}    & {\bf 0}   & \cdots & {\bf H}_{N-1} &{\bf D}\\
{\bf 0}    & {\bf 0}    & {\bf 0}   & \cdots & {\bf D}  & {\bf H}_N 
\end{array} \right ) ,
\end{equation}
which, according to Eq.~(\ref{lorenz}) has got the components:
\begin{equation}
 {\bf H}_j= \, 
  \left ( \begin{array}{ccc} 
-\sigma & \sigma & 0 \\
r-z_j & -1+D & -x_j \\
y_j & x_j & -b
\end{array} \right) 
, 
   \hspace{20mm}
 {\bf D} = \, 
  \left ( \begin{array}{ccc} 
0& 0 & 0 \\
0& -D/2 & 0 \\
0 & 0 & 0
\end{array} \right ) ,
\end{equation}
recall that $x_{(N+1)/2}=y_{(N+1)/2}=0$.
Therefore, $-b$ is always an eigenvalue with
trivial eigenvector $(0, 0, ..., 0, 1, 0, ..., 0)^T$.
Using the program {\sc matlab}
we obtained the eigenvalues for an array of $N=21$ units,
that we expect to be large enough to inform us
about the properties of a front in an infinite medium, as long as
the front is a very localized structure.

\begin{figure}
\begin{center}
\psfig{file=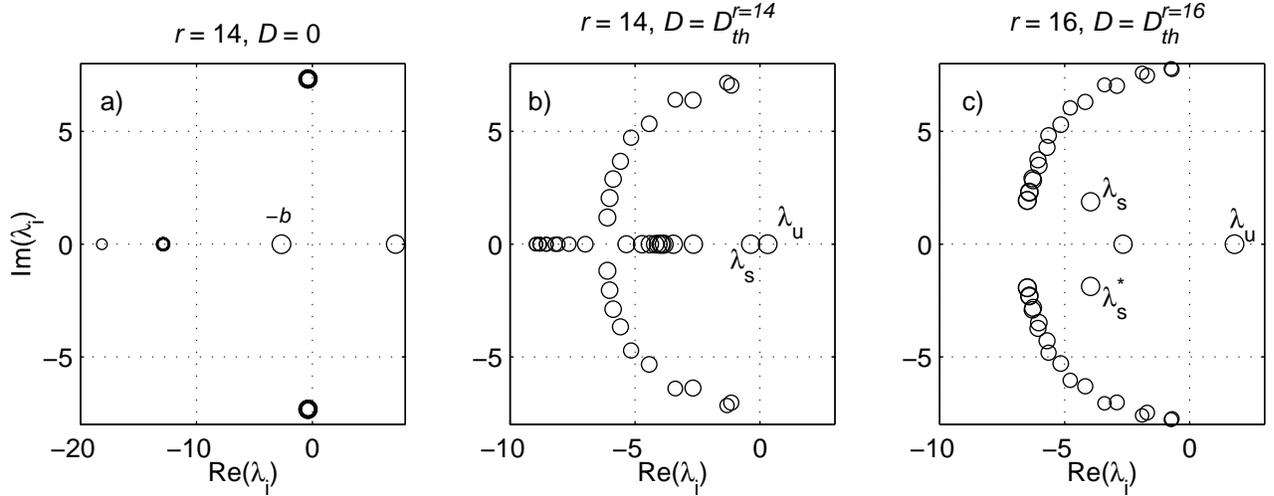}
\caption[]{Eigenvalues of the $B$-state for different 
values of $D$ and $r$.} 
\label{eigs}
\end{center} 
\end{figure}

Figure~\ref{eigs} shows the results for three values of
the parameters $r$ and $D$.  
Fig.~\ref{eigs}~(a) shows the eigenvalues 
($\lambda_i$) for $D=0$ and $r=14$,
which illustrates the meaning of the different eigenvalues.
There are three simple eigenvalues corresponding
to the oscillator located in the center of the 
array whose coordinates are 
$x_{(N+1)/2}=y_{(N+1)/2}=z_{(N+1)/2}=0$, recall
that we are considering a $B$-state.
Moreover, there are 
three ($N-1$)-degenerated eigenvalues, corresponding
to the oscillators with coordinates $C_\pm$.
Eigenvalues are 
depicted with circles of different sizes to better
observe degeneracy. 
When $D$ grows above zero degeneracy is lost.
In Fig.~\ref{eigs}~(b), we graph the eigenvalues for $D =D_{th}\approx 39.63$
(only those eigenvalues with $Re(\lambda_i)>-10$
are shown). The unstable eigenvalue
corresponds to the fixed point at the origin of a Lorenz oscillator.
There is also a slightly negative eigenvalue ($\lambda_s$) that
is the leading eigenvalue among the whole spectrum of
negative eigenvalues. 

\begin{figure}
\begin{center}
\psfig{file=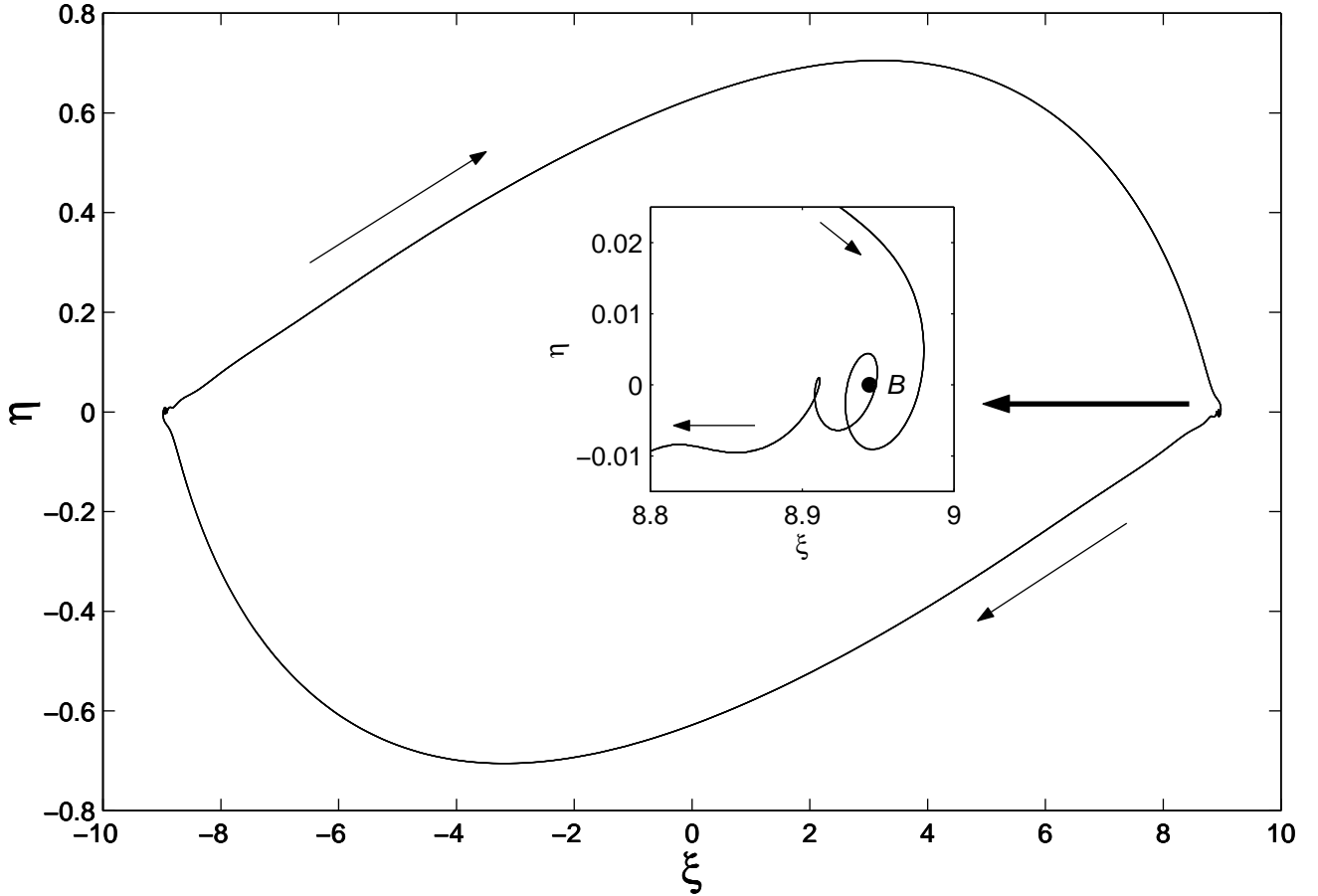}
\caption[]{Projection onto ($\xi$,$\eta$) at $r=16$, 
$D=20.136 < D_{th}\approx20.1395$. In the inset, the spiraling approach
of the trajectory to the $B$-state (filled circle) may be observed.}
\label{spi} 
\end{center} 
\end{figure}

A different scenario is found for $r=16$, as shown in Fig.~\ref{eigs}~(c).
In this case, it is not so clear which stable eigenvalue 
must be considered as the leading one in the gluing process.
There are several eigenvalues making up two lines 
at both sides of the real axis in the complex plain,
but these are not relevant, as long as they are the eigenvalues
associated with the stability properties 
of each domain at both sides of
the front where quasi-homogeneous domains around $C_{\pm}$ exist.
As well, if the eigenvalue located at $-b$ is considered, 
its eigenvector becomes trivial and it cannot   
be expected to participate in the gluing bifurcation. Then,
the solitary eigenvalue denoted by $\lambda_s (= \rho + i \omega)$
is the one (with its complex conjugate $\lambda_s^\star$) 
that defines the
leading (two-dimensional) stable manifold. In fact, 
by increasing continuously $r$ along $D_{th}$, the 
eigenvalue $\lambda_s$  could be followed directly
from that obtained for $r=14$ (see Figs.~\ref{eigs}~(b,c)). 
According to the statements 
exposed above, the graph of $\eta$ vs. $\xi$, for $D$ close to
$D_{th}$ and $r=16$, shows some 
spiraling\footnote{One should
define a new extra variable $\zeta$ to work in a 
hyper-cylindrical phase space. One could define, for instance,
$\zeta=\sum x_j y_j - b z_j$. Nonetheless, we shall 
continue to work with $\xi$ and $\eta$ only, although
keeping in mind that we are projecting the third variable.}
when the trajectory approaches the $B$-state. Figure~\ref{spi} 
shows this effect 
that confirms that now the gluing is mediated by a saddle-focus.
The cyclic definition of $\xi$ has been relaxed in order to
get a better observation of the spiral trajectory near 
the $B$-state. 
Nonetheless, if one looks at
the asymptotic properties of the oscillation period
of the front and its velocity, no difference
with the case of the planar connection is found. 
So, Fig.~\ref{r16} already confirmed the  
Eqs.~(\ref{t1}) and (\ref{t2}), and the slopes
agree with the numerical value: $\lambda_u \approx 1.7959$.

\begin{figure}
\begin{center}
\psfig{file=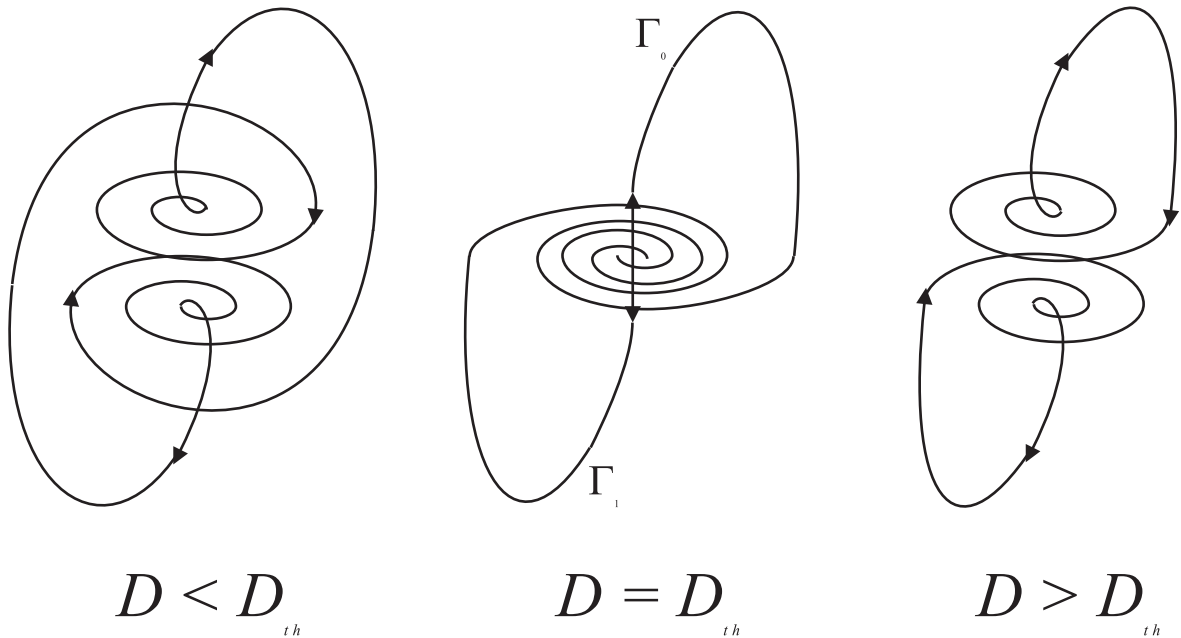}
\caption[]{Schematic of a gluing bifurcation mediated
by a saddle-focus. Two homoclinic trajectories $\it \Gamma_{0,1}$
exist at $D=D_{th}$.} 
\label{gluingsf}
\end{center} 
\end{figure}

A sketch of the gluing bifurcation
in the saddle-focus case is shown in Fig.~\ref{gluingsf}. 
Homoclinic orbits, at $D=D_{th}$ are denoted by 
$\it\Gamma_0$ and $\it\Gamma_1$. 
The process of gluing is drawn in analogy to Fig.~\ref{gluing}.
Although, from both figures, one could think that 
the complex and the real cases are almost equivalent, there
exist very fundamental differences between both cases.
Thus, the transition from oscillating to travelling
fronts may become quite convoluted in the complex case. This
is observed for larger values of $r$ 
and is the subject of the next section. 

\section{Exotic front dynamics}

\begin{figure}
\begin{center}
\psfig{file=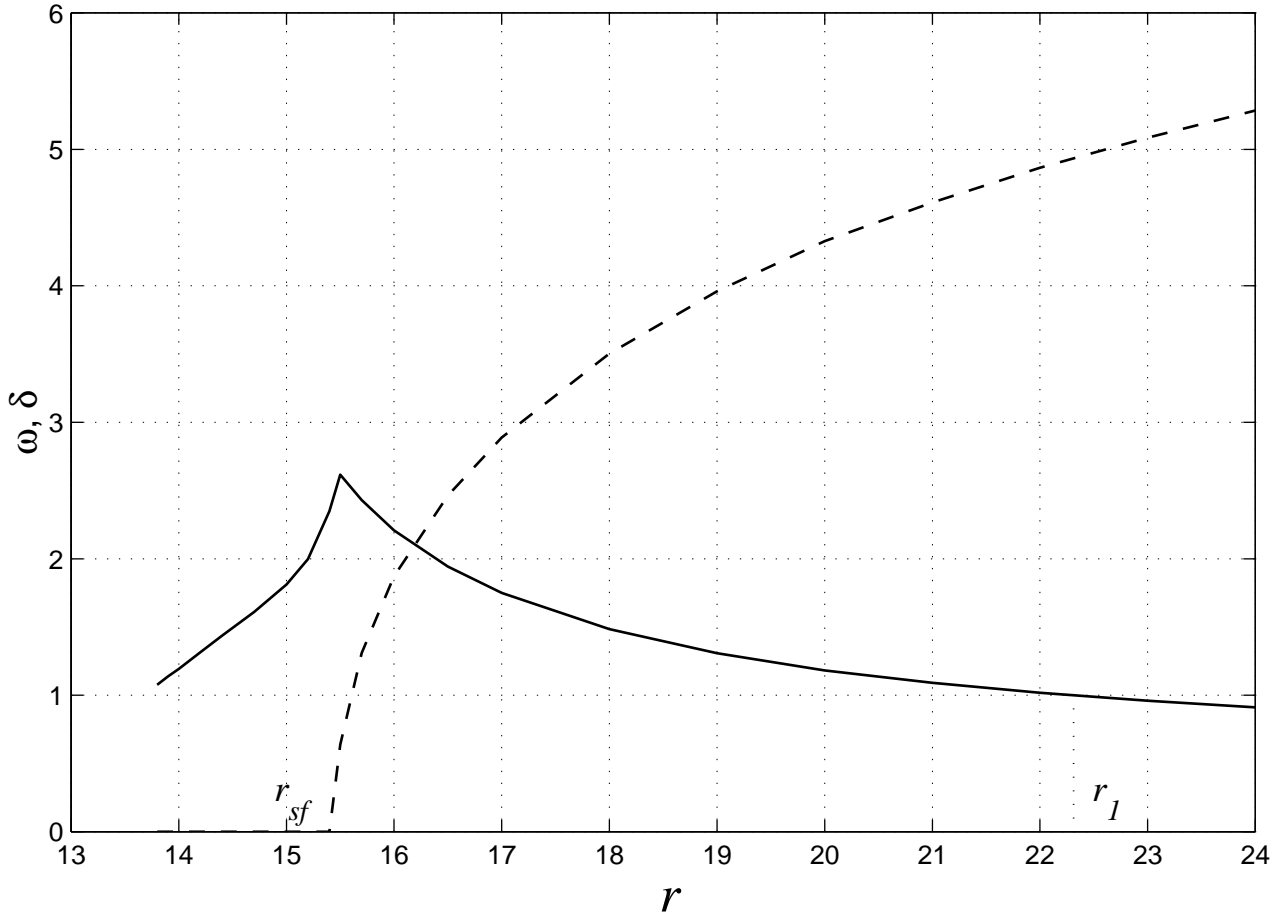}
\caption[]{Complex part of the stable leading eigenvalue $\omega$
(dashed line) of the $B$-state and saddle index $\delta$ (solid line), 
computed
along $D=D_{th}$ as a function of $r$. For $r>r_{sf}$, $\omega>0$; and
for $r>r_1$, $\delta<1$.}
\label{sindex}
\end{center} 
\end{figure}

In Figure \ref{sindex} the value of the complex part of the stable leading 
eigenvalues ($\omega$), along the line  $D=D_{th}$, 
is shown as a function of $r$ with a dashed line. It is found that 
above $r=r_{sf} \approx 15.45$ the stable leading eigenvalues
are complex conjugates, and therefore, the gluing bifurcation
occurs mediated by a saddle-focus point.
 
Many works have been devoted to the problem
of saddle-loop bifurcations involving a saddle-focus fixed point
(see e.g.~\cite{books,gucken,wiggins} and references therein).
A deep presentation of this problem is out of the scope of
this paper, so we just introduce some of the most relevant
results.

For the case of a {\it single} homoclinic orbit, occurring
at a critical value of the control parameter (e.g. $\mu=0$), it was
found by Shil'nikov~\cite{shilnikov} that when the ratio 
\begin{equation}
\delta=- {Re(\lambda_s) \over \lambda_u}=- {\rho \over \lambda_u}  \qquad,
\end{equation}
called {\it saddle index}, 
is less than one, there are a countable infinity of periodic orbits
in a neighbourhood of the homoclinic orbit, all of which are saddles.
These saddle orbits are created in a sequence of
tangent bifurcations at both sides of $\mu=0$, such that the stable 
branches (provided $\delta>1/2$) become unstable 
through period-doubling bifurcations. 
This scenario, where infinite orbits, with periods
ranging from some finite value to an infinite one,
arrange in a wiggly curve around the critical 
value, is known as {\em Shil'nikov wiggle}. And
it is, with the Lorenz and the bifocal mechanisms, 
one of the universal routes to chaos through 
homoclinicity~\cite{gucken,wiggins}.

However, the results by Shil'nikov
apply in a very small neighbourhood of the critical parameter.
Therefore, the relationship 
between the local (theoretical) results and the global 
(observed) behaviours cannot be stated oversimply.
It was shown by Glendinning and Sparrow~\cite{jsp}
that the difference between the $\delta < 1$ and the $\delta > 1$ 
approaches to homoclinicity may not be easily observed numerically 
and may not be relevant from a global point of view.
In fact, for the lowest branch (the orbit with the lowest period)
that already exist far from the critical parameter, 
deviations from the behaviour predicted by the local analysis 
are most likely to occur.

One may realize the complexity of the problem 
if one notices that, besides the Shil'nikov wiggle, 
there exist subsidiary homoclinic connections. Following
the nomenclature of Ref.~\cite{jsp}, subsidiary homoclinic
orbits are multiple-pulse homoclinic orbits, i.e. orbits
that pass several times near the stationary point without
achieving homoclinicity.

In case the homoclinic connection is {\it double} (two loops), 
as occurs in the gluing bifurcation, some differences
are appreciated with the case described above. 
The most important point is that, 
although for $\delta > 1$ no Smale horseshoe
exist as in the single case, the approach of 
orbits to ${\it \Gamma_0} \cup 
{\it \Gamma_1}$ (Fig.~\ref{gluingsf}) 
is chaotic~\cite{holmes,wiggins}.

Figure \ref{sindex} shows the value 
of $\delta$ (solid line) as a function of $r$ along 
the line $D=D_{th}$. It is found that at $r=r_{1}\approx 22.3$, 
$\delta$ crosses one. Interested in the complex
behaviours that our system could show, we focused in 
two values of $r$ above and below $r_1$: 
$r=20$ ($\delta\approx 1.18$) and $r=23$ ($\delta\approx 0.96$).
We do not want to carry out a detailed analysis 
of the solutions found, but just to show that the 
results have got the ``flavour" expected
according to previous theoretical studies
on bi-homoclinicity to a saddle-focus~\cite{holmes,glendinning,arneodo}.

\begin{figure}
\begin{center}
\psfig{file=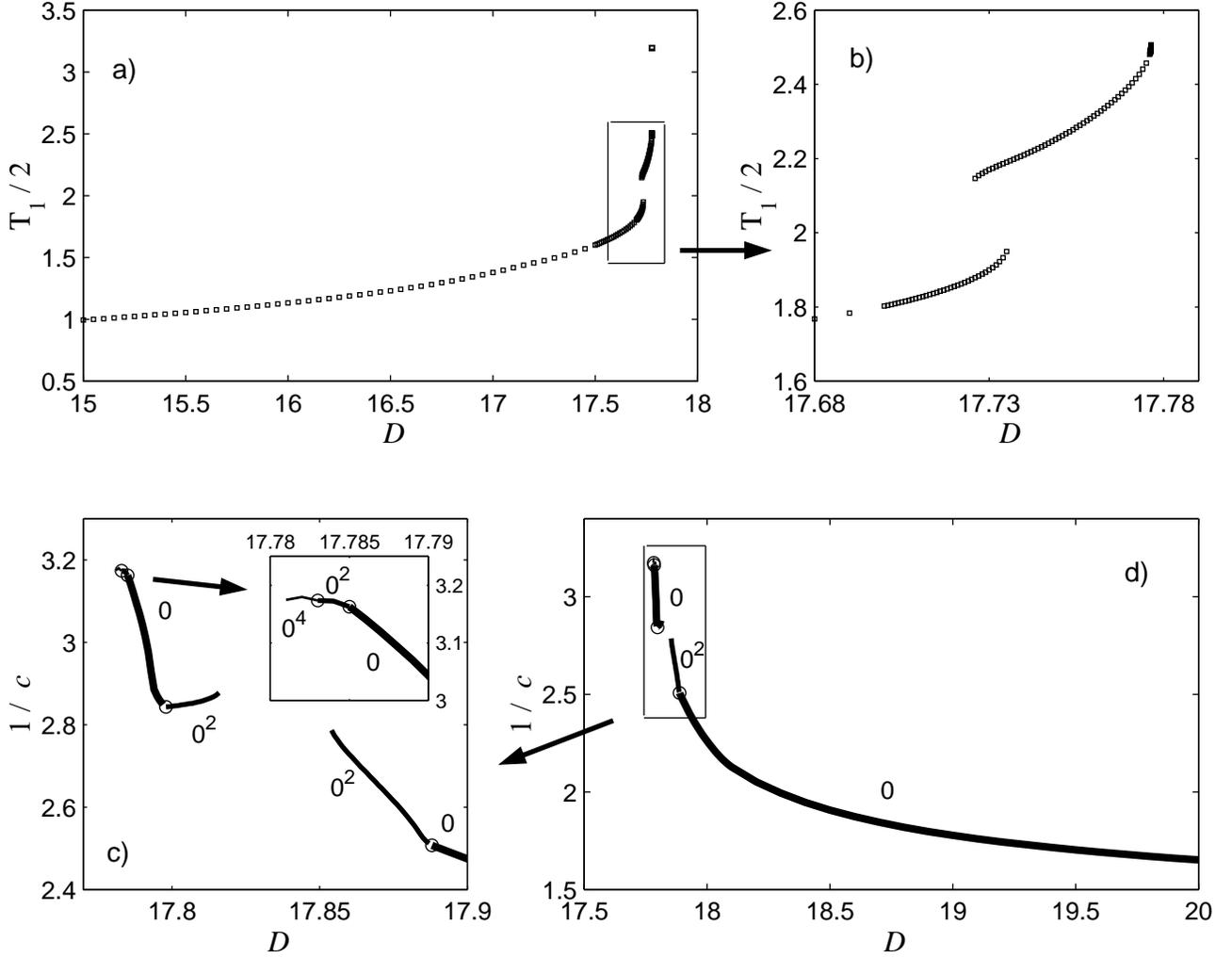}
\caption[]{$1/c$ (solid line) and $T_1/2$ (squares) as a 
function of $D$ for $r=20$. Fig. (a) 
shows that the oscillation period does not
grow continuously. An inset, showing one of this jumps
can be seen in (b). 
The velocity of the front --Fig. (d) and inset (c)-- 
is not defined in some regions where chaotic states are found.
Period-doublings are marked with circles and different
states are labelled according to its symbolic code.
Because of the symmetry all 0's may be substituted
by 1's.} 
\label{r20}
\end{center} 
\end{figure}

\subsection{$r=20$ ($\delta > 1$)}

In the following, we analyse the transition from a oscillating front
to a travelling front for $r=20$. Fig.~\ref{r20}~(a)
show the dependence of the period of oscillation 
of the front ($T_1$) vs. $D$. It is found that the period
does not grow continuously, and instead, there are some
jumps, which give rise to hysteretic cycles (Fig.~\ref{r20}~(b)
shows an inset of 12~(a)). 
These wiggles (note that unstable branches join 
consecutive stable branches) are not 
surprising according to ~\cite{jsp}, 
and are expected to appear when approaching $\delta=1$.
Unfortunately, we have not been
able to find oscillations with semi-periods larger than 3.2 t.u.
This occurs because as the period 
grows, and therefore (bi-)homoclinicity is reached, the
chaotic transients become larger and larger.

The different regimes
may be labelled by their representative symbolic sequences
of 0's and 1's, depending on 
the number of turns  at
each side of the saddle point.
For example, the uniform propagating regimes
are labeled by $\{0\}$ or $\{1\}$, but if a period-doubling 
occurs the new (stable) regimes are labeled by $\{0^2\}$ or $\{1^2\}$,
respectively. On the other hand, the oscillating 
regime, described by a two-lobed cycle, 
is represented by the code $\{01\}$ (or $\{10\}$).

For the travelling front region, the dependence
of the velocity with $D$ shows some features
that were also reported in~\cite{jsp}, where
a system of three ordinary differential 
equations, that exhibits a {\it single}
homoclinic connection to a saddle-focus,
was studied. In Fig.~\ref{r20}~ (c,d) 
the inverse of the velocity is plotted as a function of $D$. 
In those intervals of $D$ where no line appear,
the front continues to exhibit chaotic motion, 
characterized by ``spontaneous" front reversals, 
after, at least, 30000 t.u. of transient; but some small `windows'
with regular motion can be found.

\begin{figure}
\begin{center}
\psfig{file=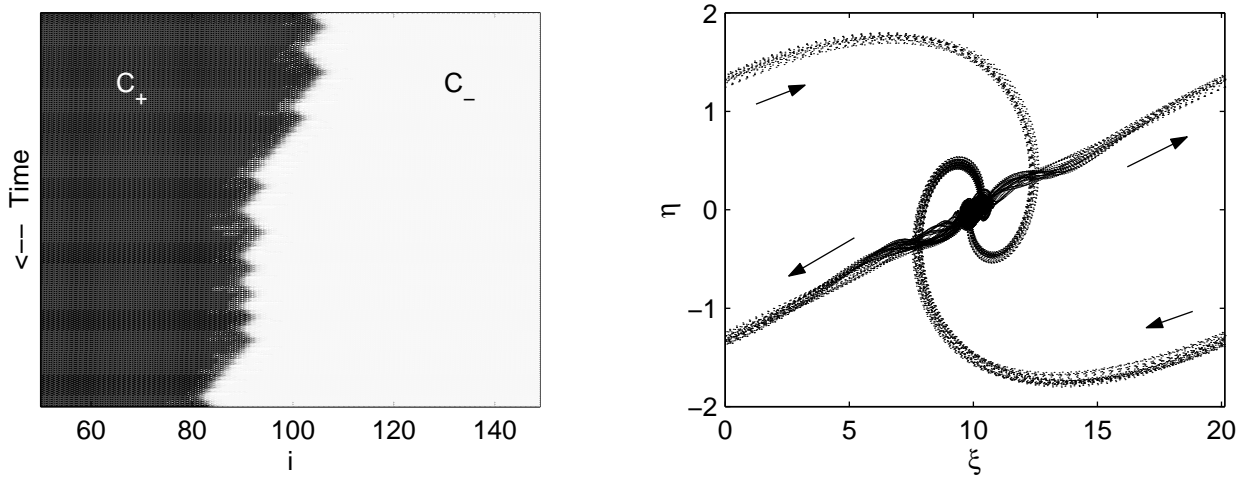}
\caption[]{Chaotic motion of the front for $r=20$
and $D=17.85$ during 200 t.u.
after a transient of 30000 t.u. a) $x$ variable 
in gray scale where abcissa corresponds to the oscillator 
number and time is running downwards.
b) Cylindrical variables $\xi$ and $\eta$ during
the 200 t.u. shown in the left figure. The interval chosen
for variable $\xi$ has been shifted half a period, with
respect to previous figures,
to better observe the homoclinic chaos
that is organized at the $B$-state.} 
\label{chaos}
\end{center} 
\end{figure}

Figure~\ref{chaos} shows the dynamics of the front
for $D=17.85$ during 200 t.u., after some transient. 
In Fig.~\ref{chaos}~(a), the variable $x$ 
of the oscillators of the array is represented in gray scale,
whereas Fig.~\ref{chaos}~(b) shows the phase portrait $(\xi,\eta)$.
The front reverses its propagation several times and
displays a quasi-erratic motion.
Although the behaviour is likely to be a chaotic transient,
from a practical point of view, it is indistinguishable
of ``true chaos". 

\begin{figure}
\begin{center}
\psfig{file=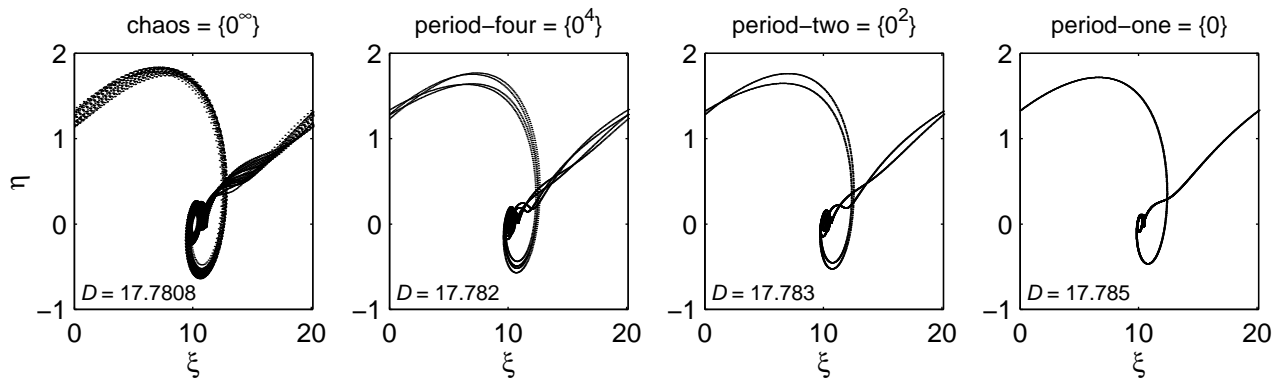}
\caption[]{From right to left
a period-doubling cascade leading to chaos.
Values of $D$ belong to the interval 
shown in the inset of  Fig.~\ref{r20} (c).} 
\label{124}
\end{center} 
\end{figure}

On the other hand, transitions between different 
regimes exhibiting sustained propagation, 
occur at period-doubling bifurcations 
(circles in Figs.~\ref{r20}~(c,d)). 
In addition, we find the well-known period-doubling
route to chaos, see the inset of Fig.~\ref{r20} (c).
In Fig.~\ref{124} three periodic orbits and a chaotic one
are shown. From right to left, a period-doubling 
cascade leads to  a type of chaos
that is characterized by sustained propagation of the front
with non-periodic velocity. This scenario,
that generates chaos with $\delta > 1$ at homoclinicity,  
was already found in~\cite{jsp}.

\subsection{$r=23$ ($\delta < 1$)}

For $r=23$, $\delta$ is less than one, and therefore 
in a small neighbourhood of $D_{th}$, one expects
to find, according to~\cite{glendinning}, Shil'nikov 
wiggles for both, the two-lobed cycle (oscillating front) and 
the one-lobed cycle (travelling front). Also we expect to find
chaos of the type reported by Arneodo {\it el al.}~\cite{arneodo};
where the strange attractor organizes around a 
saddle-focus with symmetry, like in Fig.~\ref{chaos}.

\begin{figure}
\begin{center}
\psfig{file=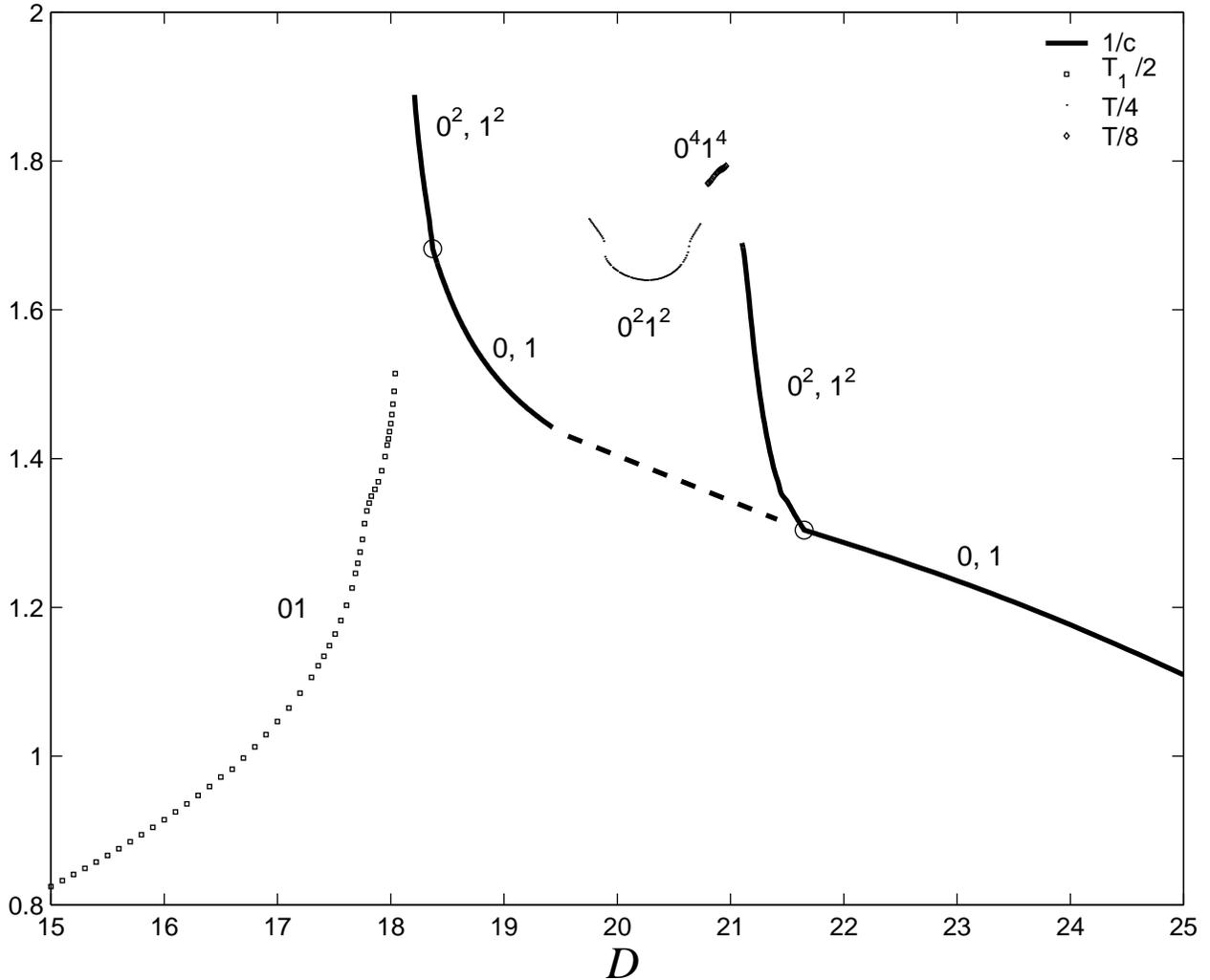}
\caption[]{$1/c$ (solid line) and $T_1/2$ (squares) as a 
function of $D$ for $r=23$. Two
periodic states with codes $0^21^2$ (dots)
and $0^41^4$ (diamonds), as well as a
conjectured unstable travelling solution
(dashed line) are shown.} 
\label{r23}
\end{center} 
\end{figure}

However, we have seen in the previous subsection,
that the local analysis is not always representative
enough of the observable results that usually
correspond to the lowest branch (i.e. the smallest period).
In fact, our results do not differ too much from those found for $r=20$.
In Fig.~\ref{r23}, oscillating, as well as, travelling solutions
are arranged according to their periods
(considering the hyper-cylindrical phase space). 

\begin{figure}
\begin{center}
\psfig{file=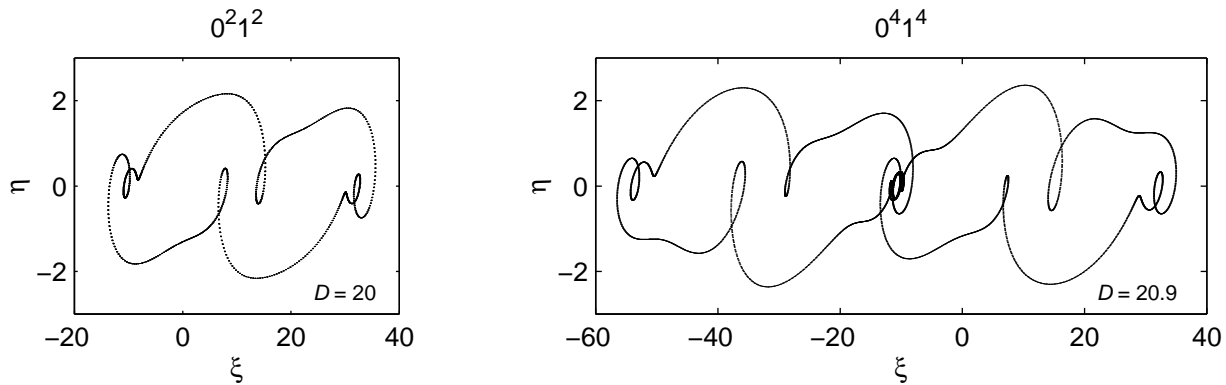}
\caption[]{Periodic orbits $0^21^2$ and $0^41^4$
for $r=23$ and two values of $D$. The 
cyclic restriction of the variable $\xi$
has been suppressed to better observe
the periodic dynamics.} 
\label{subsidiary}
\end{center} 
\end{figure}

The main oscillating solution
(code $\{01\}$) disappears in a saddle-node bifurcation at 
$D\approx 18.04$. Surely, oscillating solutions with
$\{01\}$ code and larger period exist, but we have not been able
to find them. Besides the standard oscillating solution,
we find others labelled $\{0^21^2\}$ and $\{0^41^4\}$ that
correspond to more sophisticated oscillating regimes. 
Their dynamics onto ($\xi$,$\eta$) is depicted in 
Fig.~\ref{subsidiary}. Their existence can be
attributed to subsidiary homoclinic connections.

Concerning the propagating solutions (solid line in Fig.~\ref{r23}), 
the ``period-one"
propagating solutions (codes $\{0\}$ and $\{1\}$) undergo
a period-doubling bifurcation at $D\approx 21.65$.
Nonetheless, this ``period-one" solution reappears
at $D\approx 19.43$, so we conjecture, according
to previous studies~\cite{glendinning}, that
there exist an unstable solution (dashed line) 
linking both regions. The lowest values of $D$
with propagating solution exhibit another period-doubling 
bifurcation at $D\approx18.37$.

For those intervals of $D$, where no periodic orbits were found, 
chaotic behaviour (like that of Fig.~\ref{chaos}) 
is found. Nonetheless,
some very small periodic windows are found intermingled
into the intervals with chaotic motion.

\section{Further remarks}

\subsection{Large $D$}

As the coupling $D$ is increased, a larger
amount of oscillators constitute the steepest part
of the front. In some
sense, the front becomes more `continuous' (or less steep). This
tendency allows us to understand better the 
behaviour of $D_{os}$ and $D_{th}$ in the large $D$ region.
When $D$ is large, neighbouring 
oscillators have similar $(x,y,z)$ values. Therefore,
once the $D_{os}$ line is crossed, a small increase of
$D$ (in comparison with $D_{th}$ or $D_{os}$) is needed
to achieve the multiple collision of cycles, i.e. 
the $D_{th}$ line.
As long as $D$ is large, both $A$- and $B$-states
are in a quasi-continuous and should exhibit 
similar eigenvalues spectra. 

So, as $r \rightarrow r_{\infty}\approx 13.5$,
$\lambda_u$ and $\lambda_s$ of the $B$-state
at $D_{th}$ should meet the pair of complex
conjugates eigenvalues that 
characterize the Hopf bifurcation of the $A$-state
at $D_{os}$. In short, both lines meet at $D=\infty$
in a double-zero eigenvalue (Takens-Bogdanov) point. 
For this reason, oscillations above $D_{os}$
have a small frequency and
$\delta \rightarrow 1^+$ 
when $r \rightarrow r_{\infty}$ (see Fig.~\ref{sindex}).

\subsection{The effect of parameter mismatch and asymmetry}

Our system of coupled Lorenz oscillators
is nothing but a mathematical model, 
because it is clear that it is impossible
to build an experimental set-up 
{\it infinitely large} with {\it identical} units
all of them being {\it exactly symmetric}.
This means that in a real system no 
invariance under translation and under
reflection exists.

Boundary effects can be significatively minimized 
by using an array sufficiently large, recall
that the front is very localized, and only
for large $D$ a significative amount
of oscillators are located in the steepest part 
of the front. The effect of using non-identical units
is difficult to be predicted but could
manifest in a neighbourhood of $D_{th}$
by letting some localized oscillation to exist
or making the front to reverse back at a given time. 
Nonetheless, there are lots of 
ways to make the oscillators 
non-identical and then giving rise to an 
enormous variety of effects.

More tractable is the case where 
all the oscillators are identical,
but asymmetric. In this
situation, we should use the theory
of imperfect gluing bifurcations. Larger is the
asymmetry, larger is the interval of $D$ at 
$D\approx D_{th}$
where new periodic orbits appear.
Mainly, the case when the saddle value
$\tilde\sigma=\lambda_u+ Re(\lambda_s)$ is
negative has been considered~\cite{turaev,tirapegui,gambaudo} 
showing that no more than two periodic orbits
can coexist in the same parameter domain, and
providing some simple rules 
for the symbolic sequences of 0's and 1's of
the cycles, based on a Farey tree structure.
Experimental studies can be found
in Refs.~\cite{herrero} and~\cite{imperfect}.
The latter also considers the case
with $\tilde\sigma>0$ (i.e. saddle-focus with $\delta<1$).

\subsection{$r>r_H$}

The behaviour of the system when the line
$r=r_H$ is crossed, and $C_\pm$ become unstable, 
is similar to that described for the off-diagonal 
coupling in~\cite{prejunio}. The 
transition, between oscillating
and travelling fronts, becomes
a transition between two well-differentiated 
types of spatio-temporal chaos. 
The slightly unstable nature of $C_\pm$ 
makes possible that domains with oscillators
close to $C_\pm$ exist, however, sometimes one oscillator is
able to jump to the other lobe, which,
in the propagating region, produces
the creation of two counterpropagating fronts. 
In short, for $D>D_{th}$, 
we observe spatio-temporal chaos characterized
by the spontaneous creation of counterpropagating
fronts and front reversals.

\subsection{Universality?}
\label{universality}

One may ask if the symmetry front bifurcation 
presented here is universal or not;
in other words, in which kind of system could one expect to find 
this transition?

It was shown in~\cite{predic,prejunio} that other coupling
matrices were able to induce front propagation
in an array of coupled Lorenz oscillators.
If one considers coupling matrices with all elements
zero except one, these two off-diagonal couplings:
\begin{eqnarray}
\Gamma=\gamma_{kl}=\delta_{k1}\delta_{l2} \\
\Gamma=\gamma_{kl}=\delta_{k2}\delta_{l1}
\end{eqnarray}
exhibit front propagation by a similar route 
to the one explained here. However, it may happen that
the transition occurs in such a way that the 
standing front loses its symmetry in a small interval.
This happens because a pitchfork bifurcation
renders the $A$-state unstable, and
when the coupling in increased further, 
the new (static nonsymmetric) solutions 
``collide" with the $B$-state 
transferring the stability (through a pitchfork 
bifurcation again). 
Later, it is the $B$-state
which undergoes a Hopf bifurcation, and finally, the oscillating
solution  touch the $A$-state creating the 
travelling solutions. The logarithmic laws, Eqs.~(\ref{t1}) and 
(\ref{t2}), are also obtained~\cite{predic}.

We have also checked our results with
other two bistable systems: an array whose local dynamics is
a truncation of the magnetohydrodynamic partial differential 
equations of a disc dynamo~\cite{holden}, and the 
FitzHugh-Nagumo model~\cite{fitzhugh}.

The dynamo model is similar to 
the Lorenz system and in some parameter range
it exhibits symmetric bistability. The equations
of the model are:
\begin{eqnarray}
\dot{x}&=&\alpha(y- x) \nonumber\\
\dot{y}&=&z x - y  \label{magneto}\\
\dot{z}&=&\beta- x y- \kappa z \nonumber
\end{eqnarray}
The system has one unstable fixed point at 
$P_0=(0,0,\beta/\kappa)$ and two stable
fixed points 
$P_\pm=(\pm\sqrt{\beta-\kappa},\pm\sqrt{\beta-\kappa},1)$
in the range $\beta\in(\kappa,\beta_H)$, with
$\beta_H=15$ for $\alpha=5$, $\kappa=1$.
We have found a transition like that 
with Lorenz systems, for instance for the 
coupling matrix $\Gamma=\gamma_{kl}=\delta_{k2}\delta_{l2}$.
Depending on the value of $\beta$ the transition is
smooth or chaotic. Whereas for $\beta=6$ we find
a fine logarithmic profile of the front velocity,
for $\beta=14$ the 
front velocity function is interrupted by a 
chaotic regime, like that shown in Fig.~\ref{chaos},
when approaching the threshold. 
Also the oscillating dynamics, for $\beta$ close to 
$\beta_H$ is not so simple like that shown 
in Fig.~\ref{multiple}.

For the discrete FitzHugh-Nagumo model:
\begin{eqnarray}
\dot{u}_j&=&u_j-u_j^3-v_j+D(u_{j+1}+u_{j-1}-2u_j) \nonumber\\
\dot{v}_j&=&\epsilon (u_j-a_1v_j - a_0) 
\qquad\qquad j=1,\ldots,N \label{fhn}
\end{eqnarray}
we take $a_0=0$ and $a_1=2$ which provide the ${\bf Z}_2$ symmetry
and bistability, respectively. The local dynamics 
presents a saddle equilibrium point at the origin,
and two odd symmetric stable fixed points 
$(u_{\pm},v_\pm)=(\pm \sqrt{      {{a_1-1}\over{a_1}}           },
\pm {1 \over a_1} \sqrt{  {{a_1-1}\over{a_1}}          }      )$.
As occurs in the continuous version~\cite{meron}, 
propagation only succeeds for small $\epsilon$ (${\mathcal O}(10^{-1})$).
For very small $D$, only one stable solution exits, 
the standing one ($A$-state).
When $D$ increases, two (counterpropagating) travelling solutions 
coexist with the standing one. Finally, the standing
solution undergoes a {\em subcritical} Hopf bifurcation 
and the travelling solutions become the only stable ones.
This route is apparently very different from those shown above, since
a computation of the eigenvalues corresponding to the $B$-state
reveals that $\lambda_u > -\lambda_s$ ($\delta< 1$), and therefore,
if any gluing bifurcation exists, it involves 
unstable cycles. Therefore, we speculate that
the transition is as follows:
When $D$ reaches a critical value $D_{SN}$, 
two stable travelling solutions (both propagation senses)
are born with nonzero velocity ($c=c_0\not=0$), 
in two simultaneous saddle-node bifurcations.
The unstable travelling solutions that appear in these saddle-node
bifurcations become glued, when $D$ is slightly increased, at $D=D_G$. 
In this way, the (unstable) oscillating solution, that coalesces
with the $A$-state at the (subcritical) Hopf bifurcation, is created.

\begin{figure}
\begin{center}
\psfig{file=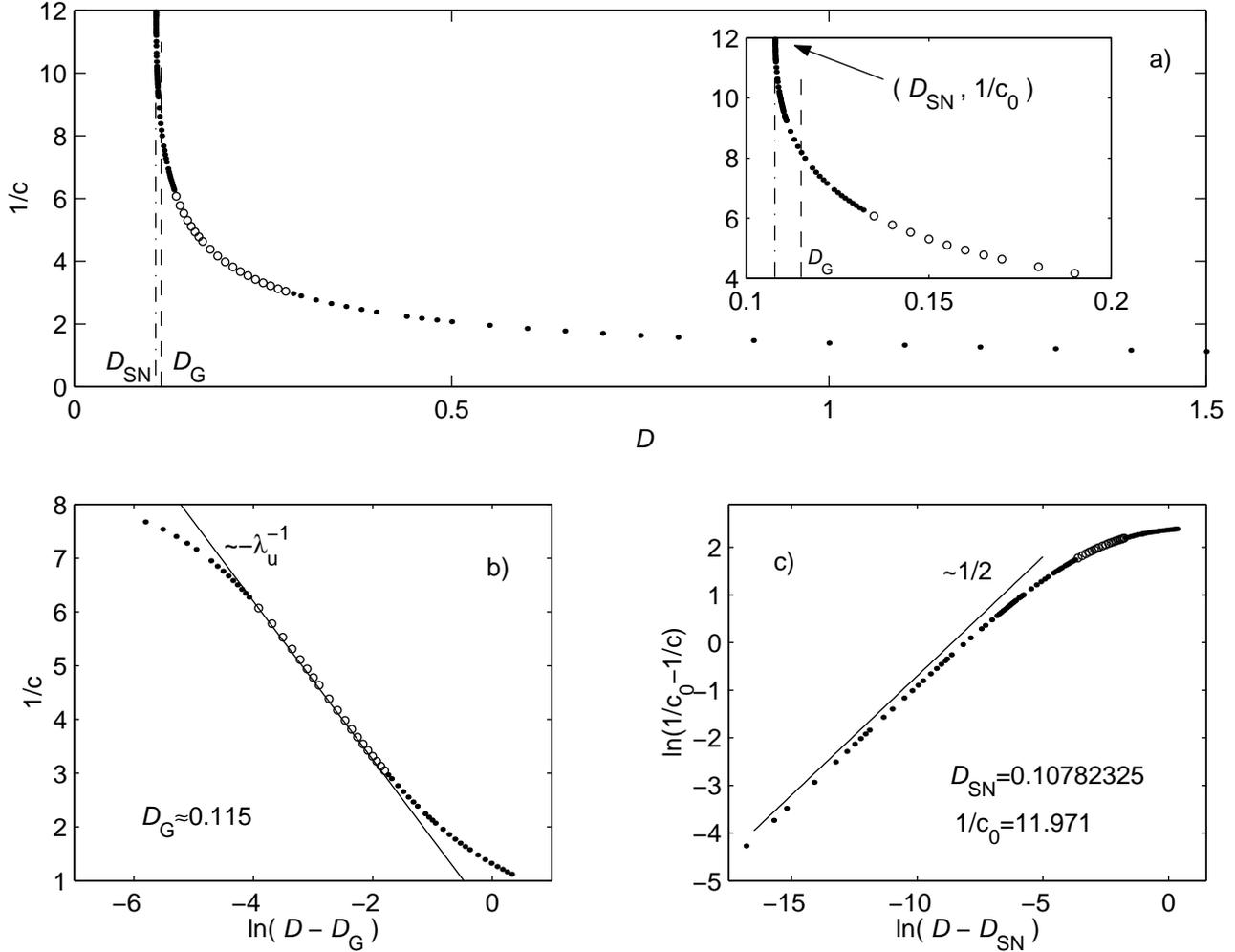}
\caption[]{Velocity of the front for the discrete FitzHugh-Nagumo
model (a). Internal parameters are $a_0$=0, $a_1=2$, 
and $\epsilon=0.1$ (see Eq.~(\ref{fhn})). Data are shown with 
dots and circles. It may be observed that 
the travelling solution ceases to exist at $D=D_{SN}$
through a saddle-node bifurcation. Nonetheless, in the 
interval shown with circles the increase of $1/c$ 
looks logarithmical. In Fig.~(b) a semilog plot 
shows that, at some interval, the slope of $1/c$ 
agrees with the value of the unstable eigenvalue 
of the $B$-solution at $D_G$: $\lambda_u=0.6781$. 
This suggests that 
the travelling solution is approaching the (non-complex) 
saddle point ($B$-state) when $D$ decreases. However 
the collision is prohibited, because the travelling solution
is stable and the saddle index of the $B$-solution
is approximately 0.7 for the values of $D$ considered,
which implies that the gluing bifurcation may 
occur between {\it unstable} cycles only. Hence,
the gluing mechanism involves the unstable travelling solutions
created at $D=D_{SN}$. In Fig.~(c) a log-log plot 
shows the dependence of $1/c$ as a function of $D-D_{SN}$.
The slope near $D_{SN}$ agrees with the expected value 0.5,
characteristic of a saddle-node bifurcation.} 
\label{fhnvel}
\end{center} 
\end{figure}

In Fig.~\ref{fhnvel}, we present the values of $1/c$
as a function of $D$, for $\epsilon=0.1$. 
One may observe that within a range of values of $D$of $D$, $1/c$
exhibits a logarithmic profile (Fig.~\ref{fhnvel}~(b))
but finally departs from that tendency, and shows a
root-square dependence, typical of a saddle-node bifurcation 
(Fig.~\ref{fhnvel}~(c)):
\begin{equation}
\left( {\frac 1 c} -{\frac 1 c_0} \right) \propto  \pm (D-D_{SN})^{1/2}.
\end{equation}
In our case, the + (resp. -) sign corresponds to the unstable (resp. stable)
solution. $D_{SN}$ is close to $D_G$ and that is the reason
why the partial logarithmic dependence can be recognized
in Fig.~\ref{fhnvel}~(b). In fact the saddle index is 
not very small, $\delta\sim0.7$, what makes possible
the unstable cycles to disappear at $D_{SN}$ close to $D_G$. 

\section{Conclusions}

It has been shown that an array composed of symmetric bistable 
units undergoes a front bifurcation originating two
counterpropagating travelling solutions. The mechanism
consists in a Hopf bifurcation of the static front, 
followed by a global bifurcation equivalent to a 
gluing bifurcation of cycles onto a cylindrical phase space. 
Accordingly, close to the threshold, the period of 
oscillation of the front ($T_1$) and 
the speed of the front ($c$)
follow logarithmic laws. An unstable
static solution mediates the gluing process, 
in such a way
that the value of its unstable eigenvalue
determines
the rate of divergence of $T_1$ and $c^{-1}$.
The transition is typically discrete, and 
the question of the continuum limit is intriguing. 

Also, it has been demonstrated that the 
gluing transition may be mediated by a saddle-focus point.
In that case, when the saddle index ($\delta$) approaches
one, the transition becomes much more convoluted.
Different oscillating and travelling regimes
are observed, included chaotic motion
of the front due to Shil'nikov chaos.

In Sec.~\ref{universality} we have dealt with other
couplings and other bistable systems. 
We have found common features with the
transition described here in detail for the Lorenz oscillator,
coupled through the $y$ variable.
Even in the case of the discrete FitzHugh-Nagumo model,
where the transition is apparently very different, a careful
examination indicates that a similar scenario exists, despite
the gluing bifurcation is ``invisible".

A further investigation is needed to know {\em a priori}
what local dynamics and what couplings are suitable 
to achieve propagation in a discrete symmetric bistable medium.
Besides the multi-variable nature of
the local dynamics, some trivial considerations 
about the coupling, and the observation that foci are more adequate 
than sinks, additional arguments are still to be developed. 
We hope that future research will clarify this point.

\section*{Acknowledgments}

The support by MCyT under Research Grant
BFM2000--0348 is gratefully acknowledged.

% \appendix

% \section{}
% \label{}

\end{document}